\documentclass[showpacs,showkeys,preprintnumbers,amsmath,amssymb]{revtex4}
\usepackage{graphicx}% Include figure files
\usepackage{bm}% bold math

\begin{document}
%\draft
\title{Laser-Cluster-Interaction in a Nanoplasma-Model with Inclusion of Lowered Ionization Energies}
\author{P. Hilse, M. Moll, and M. Schlanges}
\affiliation{Institut f\"ur Physik,
Ernst-Moritz-Arndt-Universit\"at, D-17487 Greifswald, Germany
}%
\author{Th. Bornath}
 \email{thomas.bornath@uni-rostock.de}
\affiliation{ Institut f\"ur Physik, Universit\"at Rostock,
Universit\"atsplatz 3, D-18051 Rostock, Germany
}%

\date{\today}

%------------------------------------------------------------------
\begin{abstract}
The interaction of intense laser fields with silver and argon
clusters is investigated theoretically using a modified nanoplasma
model. Single pulse and double pulse excitations are considered. The
influence of the dense cluster environment on the inner ionization
processes is studied including the lowering of the ionization
energies. There are considerable changes in the dynamics of the
laser-cluster interaction. Especially, for silver clusters, the
lowering of the ionization energies leads to increased yields of
highly charged ions.
\end{abstract}
%------------------------------------------------------------------
\pacs{52.50.Jm, 52.25Jm, 36.40.Gk, 52.38Dx}
\keywords{laser matter interaction, cluster, nanoplasma, collisional absorption, ionization}
%-------------------------------------------------------------------
\maketitle
%\keywords{laser matter interaction, cluster, nanoplasma, collisional absorption, ionization}

\section{Introduction}
The interaction of intense laser radiation with clusters is a field
of current interest. An important reason is that clusters present
the advantage of high-energy absorption. Their interaction with
electromagnetic fields is very effective because they are objects
with initially solid-like atomic density in a nanometer scale size.
Thus, absorption of energy is much larger in a cluster than for the
corresponding atomic or bulk systems at the same intensities.
Consequently, in laser-cluster experiments the emission of highly
charged ions, very energetic electrons, higher order harmonics, fast
fragments as well as strong x-rays in the multi-keV range is
observed. Different theoretical models and simulations indicate that
resonant collective absorption plays the central role. The rapid
expansion of irradiated clusters is essential as, at a certain time,
the cluster reaches the density fullfilling the resonance condition.
This can occur during a single pulse. Another method which allows a
better control is the dual-pulse laser excitation with varying time
delay between two pulses \cite{doeppner05}. Such experiments were
performed recently for silver clusters showing a strong dependence
of the highly charged ion yield as well as of the maximum energy of
emitted electrons on the delay time \cite{doeppner06}.

For the theoretical description of such dynamical processes, there
exist various methods. An overview about the different processes of
laser-cluster interation and about theoretical approaches is given,
e.g., in \cite{krainov02,reinhard_suraud,saalmann06}. More recently,
a review on theoretical concepts to describe the energy absorption by clusters in the
collisionless regime was given in \cite{popruzhenko08}.
One large
group of methods are simulations ranging from classical molecular
dynamics \cite{saalmann03,bauer05,fennel07a} used successfully for
rare gas clusters, particle-in cell methods \cite{brabec04,bauer08}
to the so-called time-dependent local-density approximation coupled
to ionic molecular dynamics \cite{reinhard01} or the
Thomas-Fermi-Vlasov molecular dynamics \cite{fennel04,fennel07b}
which treat the electrons in the cluster with density functionals
and which have been especially applied for the description of metal
clusters. In the latter method, the fermionic character of unbound
electrons is resolved, the number of active electrons in this
expensive calculations is, however, strongly limited.

Quite another method is the nanoplasma model introduced into cluster
physics by Ditmire {\it et al.} \cite{ditmire96}. In this model, the
relevant processes are described by hydrodynamic balance equations
and rate equations. For absorption and ionization rates, standard 
expressions are used. In order to understand the role of different
absorption and ionization mechanisms in large clusters, this model
seems to be suited because the influence of different approximations
for the rates can be well controlled.

An interesting topic of current discussion is the influence of the
dense cluster environment on the inner ionization processes itself.
Here, screening and ionic microfield effects play an essential role.
It has to be expected that the ionization energies of the various
ions are lowered.  There are different models whose applicability
has to be investigated thoroughly
\cite{gutierrez2005,gets_krainov06}. The entire process is not fully
understood yet, however.

The aim of the present paper is to study the influence of
correlation effects such as the lowering of the ionization energy on
the dynamics of laser-cluster interaction determining current and
internal electric field in a self consistent manner. 
In first calculations within the nanoplasma model 
\cite{bhs07_cpp,bhs07_lp} we used an improved expression for collisional 
absorption which we derived recently for the bulk plasmas case
\cite{bshk01,bshk03,hsbk05,semkat06,bs06} and
considered the effect of the external field only to avoid unphysical 
enhancement of the internal field in a self consistent treatment.
In the present paper, we follow \cite{megi2003,micheau2008} to account 
for an additional damping term corresponding to electron-surface collisions. 
The central point is that we include the lowering of the
ionization energies in the ionization rates. This lowering is
calculated using the model of Stewart and Pyatt
\cite{stewart_pyatt}.

We give a brief description of the nanoplasma model and its
modifications in Section 2. The consequences of lowered ionization
energies on the dynamics of the interaction of silver and argon
clusters with intense laser fields are discussed in Section 3 for
the cases of single pulse excitation as well as for dual pulse
excitation.

\section{The nanoplasma model}
% ##################################################
The nanoplasma model allows to describe different physical processes
like ionization, heating, and expansion that occur during the
laser-matter interaction. In this section, a short review of the
model is given and some generalizations are discussed. Especially,
the influence of screening on the ionization rate will be
considered. A further modification concernes the calculation of the
heating due to collisional absorption.

The clusters are assumed to be initially
neutral spheres with uniform temperature and density.
It is assumed that there will be no gradients in density and
temperature during the further evolution of the nanoplasma.
The plasma will be heated due to the interaction with the laser radiation which is described
by collisional absorption (inverse bremsstrahlung).
The expansion and the resulting cooling of the plasma is modeled
with hydrodynamic equations. For all processes, appropriate
approximations have to be found to build up a system of rate
equations for the plasma composition and a  set of hydrodynamic
equations in order to simulate the dynamics of the plasma.
\subsection{Ionization}
In the model, it is assumed that all electrons are bound before the
cluster is irradiated with the laser beam. For tunnel ionization,
cycle-averaged rates from Ammosov, Delone, and Krainov (ADK)
\cite{adk} where used applied for the internal electric field of the
cluster (see Section 2). The second important ionization process is
electron impact ionization. Here one can distinguish ionization due
to collisions with thermal electrons and ionization processes due to
the directed oscillating motion of the electrons in the laser field
\cite{ditmire96}. The ionization rates for the various ionization
stages are based on the semi-empirical cross sections of Lotz
\cite{lotz}, for details, see also \cite{bhs07_lp}.

In a dense system, scattering processes are influenced by the
surrounding medium. However, the evaluation  of cross sections in
strongly coupled systems is a demanding task. Undoubtly, the
lowering of ionization energies is a main nonideality effect in
dense Coulomb systems \cite{bluebook}. Here we  use for the ionization rates due
to impact of thermal electrons
the simple relation \cite{sbk88,bhs07_lp}
\begin{equation}\label{nonidealrate}
 W_{Z}=W^0_{Z} \exp{\left(-\frac{\Delta_{Z}}{k_BT}\right)}
\end{equation}
with the ideal rates $W^0_{Z}$ calculated with the ionization cross
sections for the isolated scattering event. $\Delta_{Z}$ is the
shift of the ionization energy. According to (\ref{nonidealrate})
there is an exponential increase of the ionization rate with
decreasing ionization energy.

For the lowering of ionization energy due to screening, Stewart and
Pyatt \cite{stewart_pyatt} derived a formula interpolating between
the cases of Debye screening and of the ion-sphere model. Consider
the ionization reaction $A_{Z-1} \longrightarrow A_{Z} + e^-$. In a
plasma medium, the ionization energy $I_{Z-1}=I^0_{Z-1} +
\Delta_{Z-1}$ is lowered due to screening. Stewart and Pyatt have
calculated the lowering of ionization potentials in plasmas beyond
the Debye screening case. Starting from the finite-temperature
Thomas-Fermi model, they evaluated the effect of the free electrons
and the neighboring ions on the potential distribution around an
ion. The following approximate formula could be derived:
\begin{equation}\label{sp_shift}
\Delta_{Z-1}= -\frac{[3(Z^*+1)\frac{Z\kappa
e^2}{k_BT}+1]^{2/3}-1}{2(Z^*+1)}\,k_BT
\end{equation}
where $\kappa$ is an effective inverse screening length
\begin{equation}
\kappa^2 = \frac{4\pi e^2}{k_B T}(Z^* +1)n_e
\end{equation}
with the free electron density $n_e$. $Z^*$ is an effective charge number
\begin{equation}
Z^*=\frac{\langle Z^2 \rangle}{\langle Z \rangle}\quad \mbox{with}
\quad
\langle Z^k\rangle =\frac{\sum_{Z=0}^{Z_{\rm max}}{n_Z Z^k}}{\sum_{Z=0}^{Z_{\rm
max}}{ n_Z}}
\end{equation}
where $n_Z$ is the particle number density of ions with charge $Z$,
$Z_{\rm max}$ denotes the fully stripped ion.

Gets and Krainov have calculated recently \cite{gets_krainov06} the ionization potentials for
ions in rare gas clusters using the Schr\"odinger equation with the Debye potential.
In a first-order perturbation approach using Coulomb matrix elements, they derived a shift which
agrees well with numerical solutions of the Schr\"odinger equation.
The Debye shift -- which is the lowest order -- overestimates the shift slightly.
For stronger coupling, low temperature or high density, the calculated shift of
Gets and Krainov is bigger than the Stewart-Pyatt shift. By using the Stewart-Pyatt model, it
is possible to extend the description to parameter regions where
the condition of linear screening is fulfilled no longer.

%##############################################################################################

\subsection{Plasma heating by collisional absorption}

The balance equation for the electrical current density in a spherical cluster can be
written in the following form
\begin{eqnarray}
\frac{{\rm d} }{{\rm d}t}{\bf j}(t) - \frac{\omega^2_p}{4\pi}\,{\bf E}^{\rm ext}(t)
+\frac{\omega^2_p}{3} \int_{t_0}^t {\rm d}{\bar t}\,{\bf j}({\bar t})
=\sum_c
\int \frac{d^3 p_c}{(2\pi\hbar)^3} \frac{e_c{\bf p}_c}{m_c} I_c({\bf p}_c,t)
\end{eqnarray}
with the plasma frequency $\omega^2_p=\sum_c  \frac{4\pi e^2_c
n_c}{m_c}$ and $I_c$ being the general collision integral for
species $c$. The third term on the left hand side describes the
polarization stemming from the mean field contribution. The right
hand side of the above equation describes the friction $-{\bf R}$
due to collisions. The balance equation then can be written as
\begin{eqnarray}\label{balance}
\frac{{\rm d} }{{\rm d}t}{\bf j}(t) + {\bf R}\{{\bf j}\}&=&
\frac{\omega^2_p}{4\pi}\,{\bf E}^{\rm ext}(t) -\frac{\omega^2_p}{3}
\int_{t_0}^t {\rm d}{\bar t}\,{\bf j}({\bar t})\\
&=&\frac{\omega^2_p}{4\pi}\,{\bf E}^{\rm int}(t) \,,\nonumber
\end{eqnarray}
where the internal electrical field was introduce in the second
line. The friction term is a nonlinear non-Markovian functional of
the current, ${\bf j}=\sum_c n_c e_c {\bf v}_c $,  for details, see
\cite{bshk03,bluebook,bs06,bhs07_lp}. 
Considering harmonic electrical fields, ${\bf E}={\bf E}_0 \cos{(\omega t + \phi)}$,
the component $R(\omega)$ can formally be written as
$R(\omega)=\nu(\omega,j(\omega))\,j(\omega)$, where the complex
collision frequency, however, is dependent on the current. The
balance equation takes the form
\begin{eqnarray}\label{strombilanz}
-i\omega\,j(\omega) + \nu(\omega)\,j(\omega)=
\frac{\omega^2_p}{4\pi}\,E^{\rm int}(\omega)
\,,
\end{eqnarray}
where $E^{\rm int}(\omega)$ is the Fourier component of the internal
field. We can introduce the field dependent(nonlinear) conductivity
$\sigma$ according to
\begin{eqnarray}
j(\omega)=\sigma(\omega)\,E^{\rm int}(\omega)=
\frac{\frac{\omega^2_p}{4\pi}\,}{-i\omega+\nu(\omega)}\,E^{\rm
int}(\omega) \,.
\end{eqnarray}
The quiver velocity ${v}_{\rm os}$ is given then by
\begin{eqnarray}\label{quiver}
{v}_{\rm os}=\frac{|j(\omega)|}{e\,n_e}= \frac{e|E^{\rm
int}(\omega)|}{m_e|-i\omega+\nu(\omega)|} \,.
\end{eqnarray}
Approximately, one can use the familiar relation ${v}_{\rm
os}\approx eE^{\rm int}/(m_e\omega)$.

External and effective field are connected by
\begin{eqnarray}\label{E-in}
E^{\rm int}(\omega)&=&
\frac{\omega+i\nu(\omega)}{\left(\omega-\frac{\omega^2_p}{3\omega}\right)+i\nu(\omega)}\,E^{\rm ext}(\omega)\\
&=&\frac{3}{2+\varepsilon(\omega)}\,E^{\rm ext}(\omega) \,,
\nonumber
\end{eqnarray}
where we introduced the (macroscopic) dielectric function
$\varepsilon(\omega)$ of the system which is connected to the
internal conductivity by
$\varepsilon(\omega)=1+\frac{i4\pi\sigma(\omega)}{\omega}$.

The energy absoption rate is given by
\begin{eqnarray}
\frac{\langle {\bf j}\cdot{\bf E}^{\rm ext}\rangle}{\frac{1}{4\pi}\langle {\bf E}^{\rm ext}\cdot{\bf E}^{\rm ext}\rangle}
&=&\frac{\omega_p^2}
{\left[\omega-\frac{\omega^2_p}{3\omega}-{\rm Im}\nu(\omega)\right]^2+\left[{\rm Re}\nu(\omega)\right]^2}
\,{\rm Re}\nu(\omega)
\,.\nonumber
\end{eqnarray}
Here the brackets denote cycle-averaged quantities
\cite{bshk01,bshk04}. In particular, we have $\langle {\bf E}^{\rm
ext}\cdot{\bf E}^{\rm ext}\rangle=|E_0|^2/2$ with $E_0$ being the
amplitude of the external field. Often the imaginary part can be
neglected, and we get for ${\rm d} U/{\rm d}t=\langle{\bf
j}\cdot{\bf E}^{\rm ext}\rangle$
\begin{eqnarray}\label{eintrag}
\frac{{\rm d} U}{{\rm d} t} &\approx&\frac{\omega_p^2\, {\rm
Re}\nu(\omega)}{8\pi} \frac{1}
{\left[\omega-\frac{\omega^2_p}{3\omega}\right]^2+\left[{\rm
Re}\nu(\omega)\right]^2} \,  |E_0|^2 \,,
\end{eqnarray}
which is, setting ${\rm Re}\nu(\omega)=\nu$ (as usual),
the same expression for the heating rate as in \cite{ditmire96}.

The dynamical collision frequency is dependent on the frequency of the field
and, in general, also on the field strength. That's why the inner electrical 
field and the collision frequency have to be determined in a self consistent manner, see eqs. (\ref{quiver}) and
(\ref{E-in}). 
Taking into account only electron-ion collisions, cf. eq. (\ref{collfreq}), the inner 
electrical field is enhanced in the vicinity of the Mie resonance condition,
$\omega\approx \omega_p/\sqrt{3}$, unphysically strong. There are different attempts to
avoid this behavior, e.g., \cite{liu2001,megi2003}. Popruzhenko et al. \cite{popruzhenko08}
have given a thorough discussion of such mechanisms in terms of collisionless absorption.
Here we follow \cite{megi2003,micheau2008},
and consider
not only collisions of electrons with individual ions but also with the surface of the cluster. 
Therefore, we will use 
\begin{eqnarray}\label{nu}
\nu(\omega) ={\rm Re}\nu_{ei}(\omega)+\nu_{s}(\omega) \,,
\end{eqnarray}
For the electron-ion collision frequency $\nu_{ei}$, an expression is used which has been
derived in a quantum statistical approach for the case of bulk plasmas \cite{bshk03,bluebook,bs06}.
\begin{equation}
\label{collfreq} {\rm Re}\nu_{ei}=\frac{16\pi n_i Z^2
e^4}{\omega_p^2 m_e^2 {v}_{\rm os}^2}\int \frac{d^3q}{(2\pi\hbar)^3}
\sum_{m=1}^\infty m\omega J_m^2\left(\frac{{\bf q}\cdot{\bf v}_{\rm
os}}{\hbar\omega}\right)V(q) {\cal S}_{ii}({\bf q}) {\rm
Im}\varepsilon^{-1} ({\bf q},-m\omega)
\end{equation}
where $n_i$ and $e_i$ denote the ion density and charge, $J_m$ are
Bessel functions of the first kind, $V(q)=4\pi\hbar^2/q^2$, ${\cal
S}_{ii}$ denotes the static ion-ion structure factor, and
$\varepsilon$ is the Lindhard dielectric function for the electrons.
This expression does not involve a Coulomb logarithm. The quantum
statistical treatment \cite{bshk01,kull01} automatically leads to
convergent expressions for the collision frequency for higher
coupling where the classical results break down because some cut-off
is introduced \cite{decker} leading to the Coulomb logarithm. Strong
correlations of the ionic subsystem are expressed by the ionic
structure factor. Cold ions decrease the collision frequency in a
wide range of the plasma-parameters \cite{hsbk05,semkat06} what was
also confirmed by molecular dynamics simulations \cite{hsbk05}.

The electron-surface collision frequency is taken as \cite{megi2003,micheau2008}
\begin{equation}
\label{surface} \nu_{s}=\frac{\sqrt{{v}^2_{\rm th}+{v}^2_{\rm os}}}{r}
\end{equation}
with the cluster radius $r$, the thermal velocity ${v}_{\rm th}$ and
the quiver velocity ${v}_{\rm os}$.

\subsection{Cluster expansion}
After formation of the nanoplasma, the clusters are expanding during and after  the laser
pulse mainly due to the pressure of the hot electrons and a possible charge buildup at the
cluster. Heating of the plasma via inverse bremsstrahlung as well as the production of
electrons in ionization processes lead to an increasing electron pressure
with the consequence that the plasma will expand in the surrounding vacuum. This expansion
is governed by the following equation \cite{reinhard_suraud} (please note, that there is a misprint
in eq. (27) in \cite{ditmire96})
\begin{equation}
 \frac{{\rm d}^2r}{{\rm d}t^2}=5\, \frac{p_e+ p_{\rm coul}}{n_i
 m_i}\,\frac{1}{r}
\end{equation}
where $r$ denotes the plasma radius, and $n_i, m_i$ are the ion
density and mass. The pressure is taken here as the ideal electron
Fermi gas pressure. The Coulomb pressure contribution
$p_{\rm coul}$ due to a charge buildup \cite{ditmire96} is comparatively
small in the calculations following below.

The expansion of the cluster leads to a cooling. The equation of motion for the temperature is
\begin{equation}
\frac{{\rm d}T}{{\rm d}t}=-2\frac{T}{r} \frac{{\rm d}r}{{\rm
d}t}+\frac{1}{c}  \frac{{\rm d}U}{{\rm d}t}.
\end{equation}
Here $ \frac{{\rm
d}U}{{\rm d}t}$ is the expression (\ref{eintrag}) for the energy impact and $c=3/2n_ek_B$ is
the heat capacity of an ideal gas.

\section{Numerical results}
First, we adopt single lase pulses with a Gaussian shape in the
intensity profile and a full width at half of the maximum (FWHM) of
130fs. The wavelength of the laser is choosen to be 825nm. The Mie
resonance can occur therefore at an electron density $n_e=3n_{\rm
crit}$ with a critical plasma density $n_{\rm crit}=1.64 \times
10^{21} {\rm cm}^{-3}$. Most of the calculations are for silver
clusters. Furthermore, argon clusters are considered.
%###################################################################################
\begin{figure}[htb]
\includegraphics[width=0.42\textwidth,angle=270]{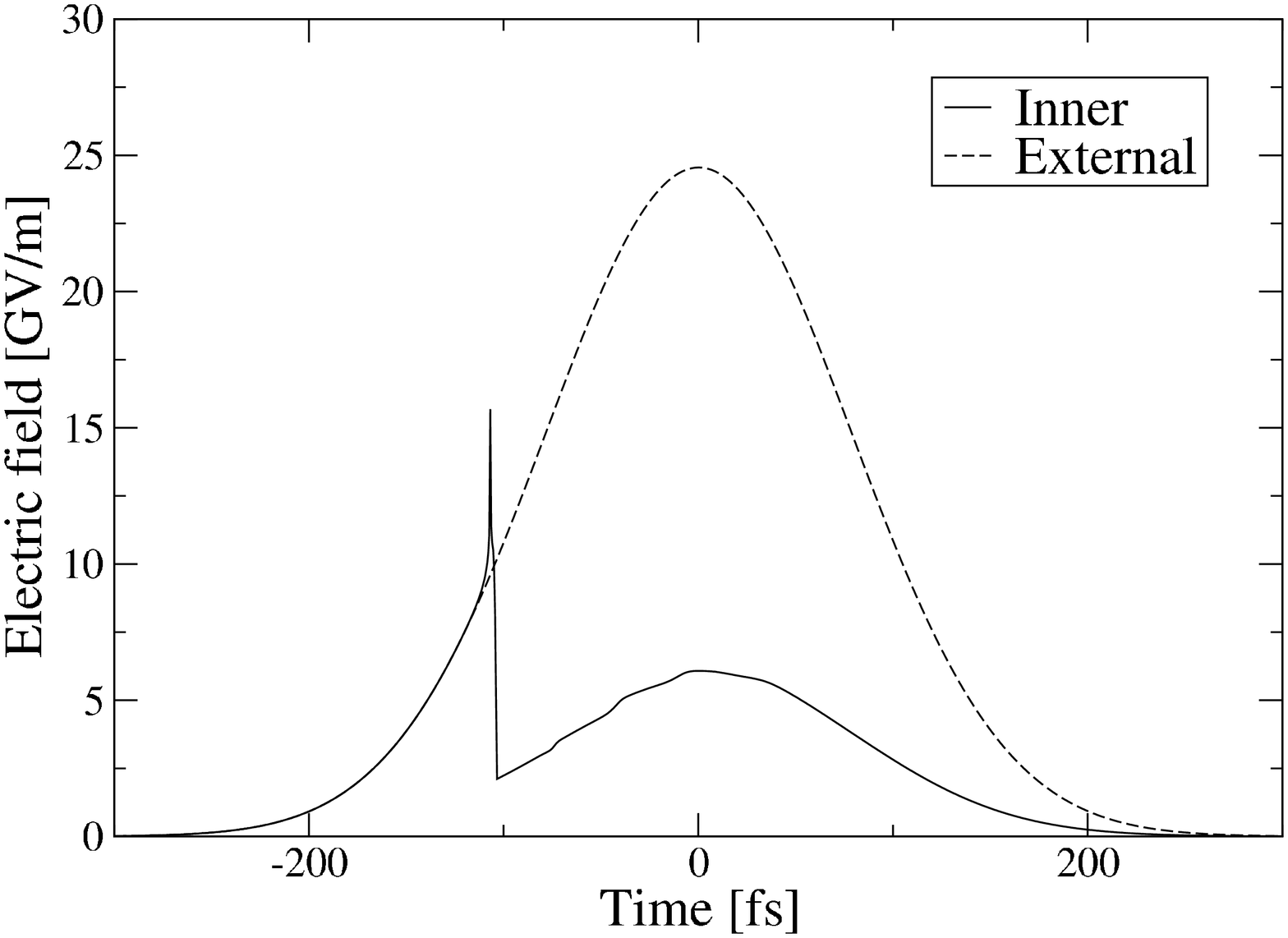}
\caption{{External and inner electrical fields in a silver
nanoplasma with initial cluster diameter of 60nm. The 130 fs laser
pulse at 825 nm has a peak intensity of 80 TW${\rm cm}^{-2}$.
\label{fields}}}
\end{figure}
%###################################################################################

As already mentioned above, the nanoplasma model represents a
coupled set of hydrodynamic and rate equations, which allows a
description of the dynamics of the interaction of intense laser
fields with clusters. In particular, the temporal evolution of,
e.g.,  the density, the cluster radius, the electron temperature, and
the occupation numbers of the different ionic charge states can be
calculated. In the present calculations, special attention is paid
to the influence of the lowering of the ionization energy on
dynamics of the laser-cluster interaction. Concerning the heating
and ionization rates, eq.~(\ref{balance}) cannot be solved simply
for $j$ in the general case. Current and internal field have to be
determined in a self consistent manner according to
eq.~(\ref{quiver}) and eq.~(\ref{E-in}) with a collision frequency
$\nu$ depending on the internal field via the quiver velocity
${v}_{\rm os}$ in a nonlinear way.

\begin{figure}[htb]
\begin{minipage}[t]{.45\textwidth}
\includegraphics[width=0.9\textwidth,angle=270]{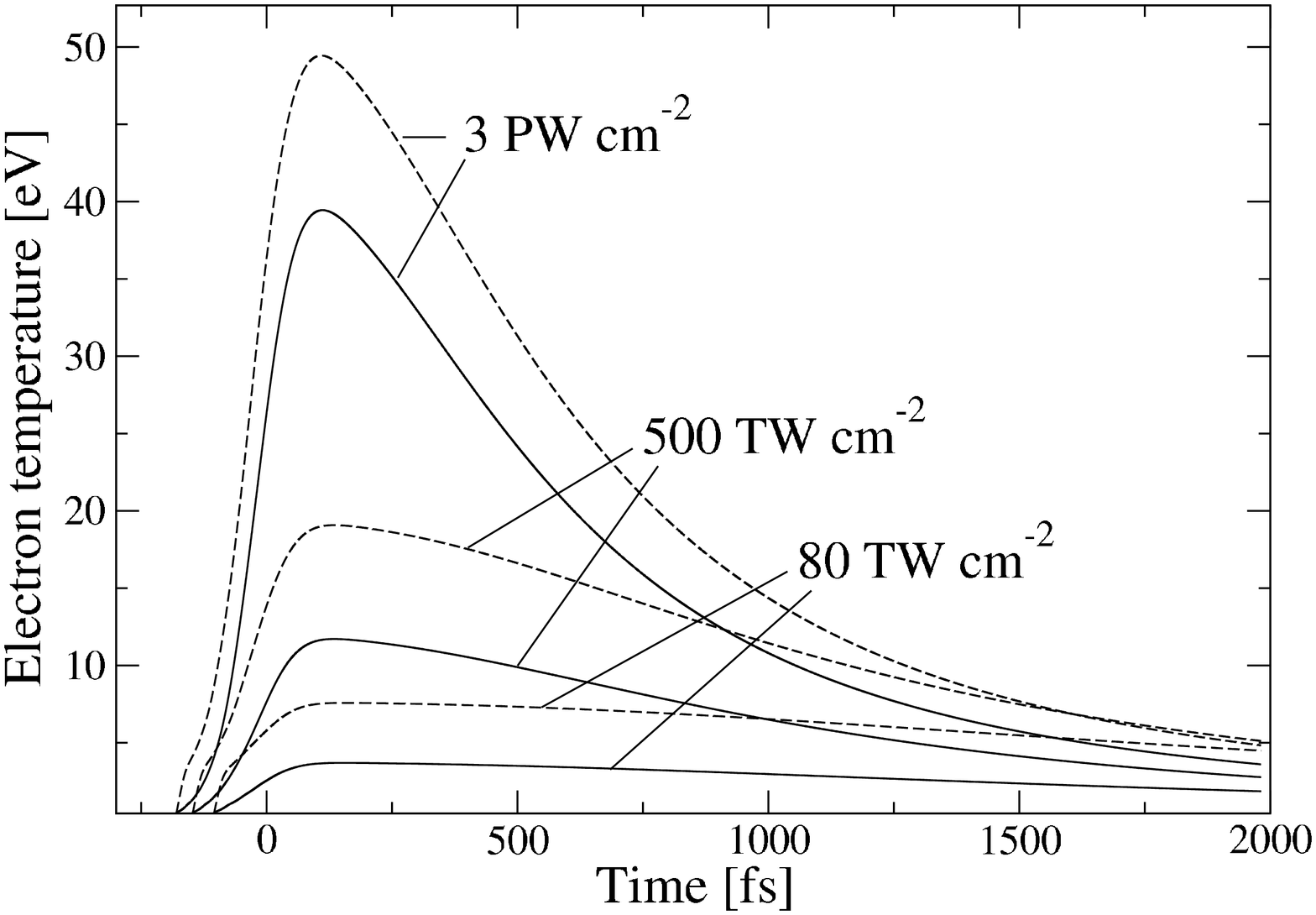}
\caption{{Electron temperature as a function of time for a silver
nanoplasma with initial cluster diameter of 60nm for different
intensities of the 130 fs pulse at 825 nm. Solid line: Lowering of
ionization energies with Stewart-Pyatt shift. Dashed line: No shift.
\label{temp_silber_single}}}
\end{minipage}
\hfil
\begin{minipage}[t]{.45\textwidth}
\includegraphics[width=0.9\textwidth,angle=270]{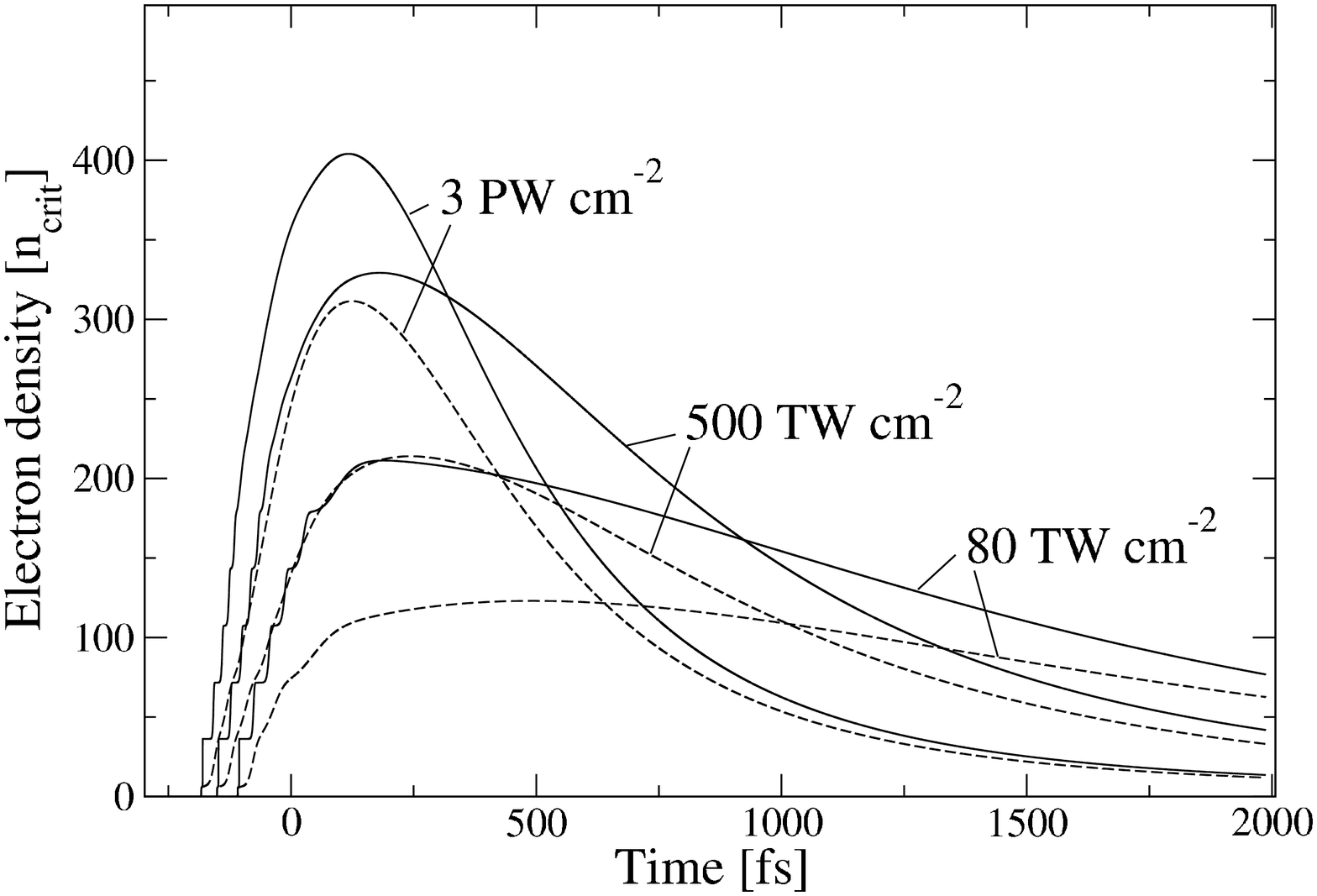}
\caption{{Electron density as a function of time for a silver
nanoplasma with initial cluster diameter of 60nm for different
intensities. The pulse parameters are the same as in
Fig.~\ref{temp_silber_single}. Solid line: Lowering of ionization
energies with Stewart-Pyatt shift. Dashed line: No shift.
\label{dichte_silber_single}}}
\end{minipage}
\end{figure}

Results for the internal field in a silver cluster with an initial
diameter of 60 nm irradiated by 130 fs laser pulse with a peak
intensity of 80 TW${\rm cm}^{-2}$ are shown in Fig.~\ref{fields}.
Free charges are produced in the beginning of the laser-cluster
interaction by tunnel ionization. The produced nanoplasma is then
heated by collisional absorption, and a further increase of the
densities of free charges is mainly determined by the electron
impact ionization processes. The internal field increases up to the
Mie resonance which is reached for an electron density $n_e=3n_{\rm
crit}$. There, an enhanced peak occurs followed by a sharp decrease
of the internal field. After the resonance the electron density
remains overcritical up to the end of the laser pulse (see also
Fig.~\ref{dichte_silber_single}) leading to values of the internal
field considerably lower than those of the external one.

%###################################################################################
%###################################################################################
%###################################################################################
\begin{figure}[htb]
\begin{minipage}[t]{.45\textwidth}
\includegraphics[width=0.9\textwidth,angle=270]{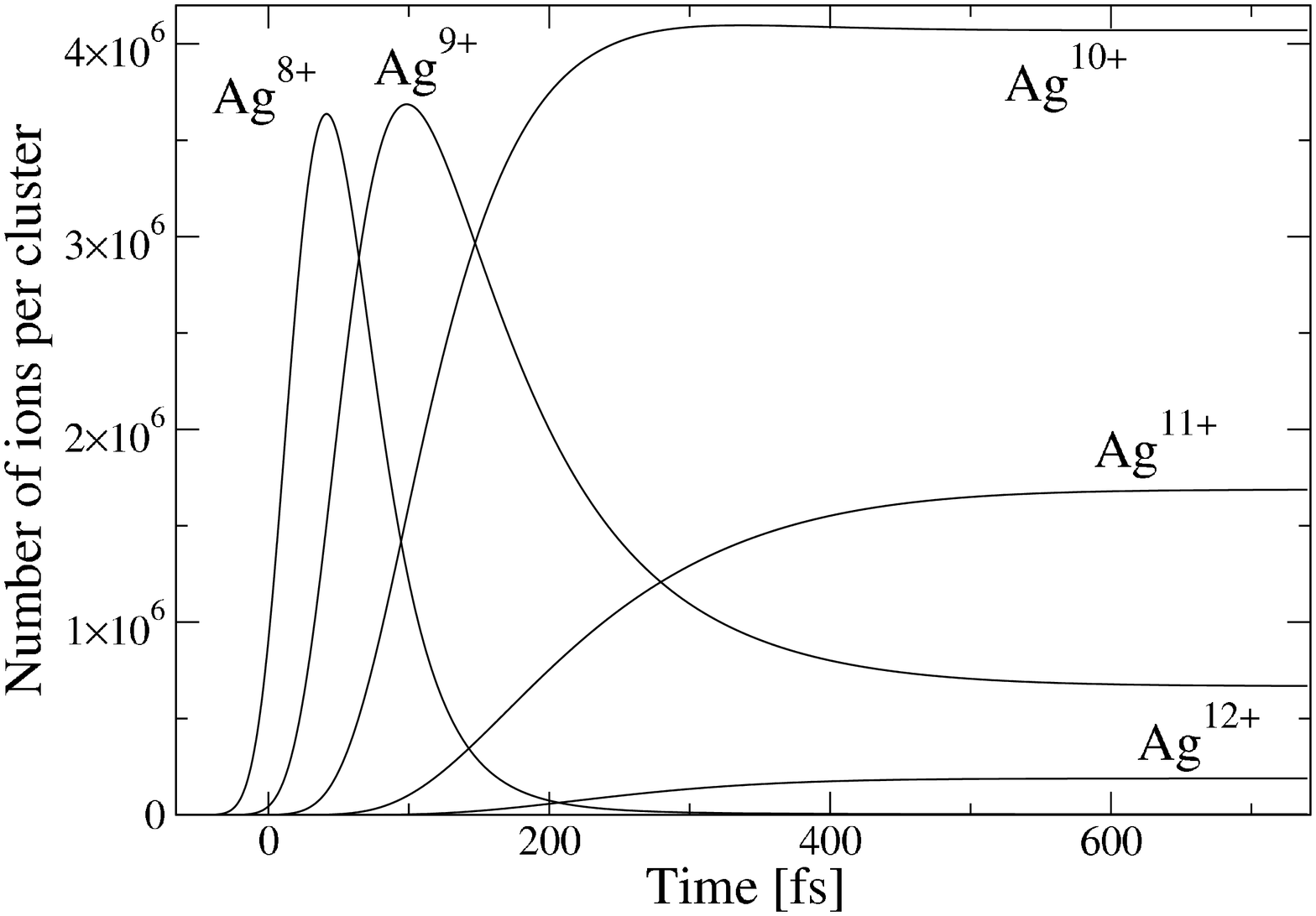}
\caption{{Composition as a function of time for a silver nanoplasma
with initial cluster diameter of 60nm. There was no shift of
ionization energies considered in this calculation. The laser pulse
is Gaussian with 130fs length and a peak intensity of 3PW${\rm
cm}^{-2}$. \label{comp_silber_single_o_shift}}}
\end{minipage}
\hfil
\begin{minipage}[t]{.45\textwidth}
\includegraphics[width=0.9\textwidth,angle=270]{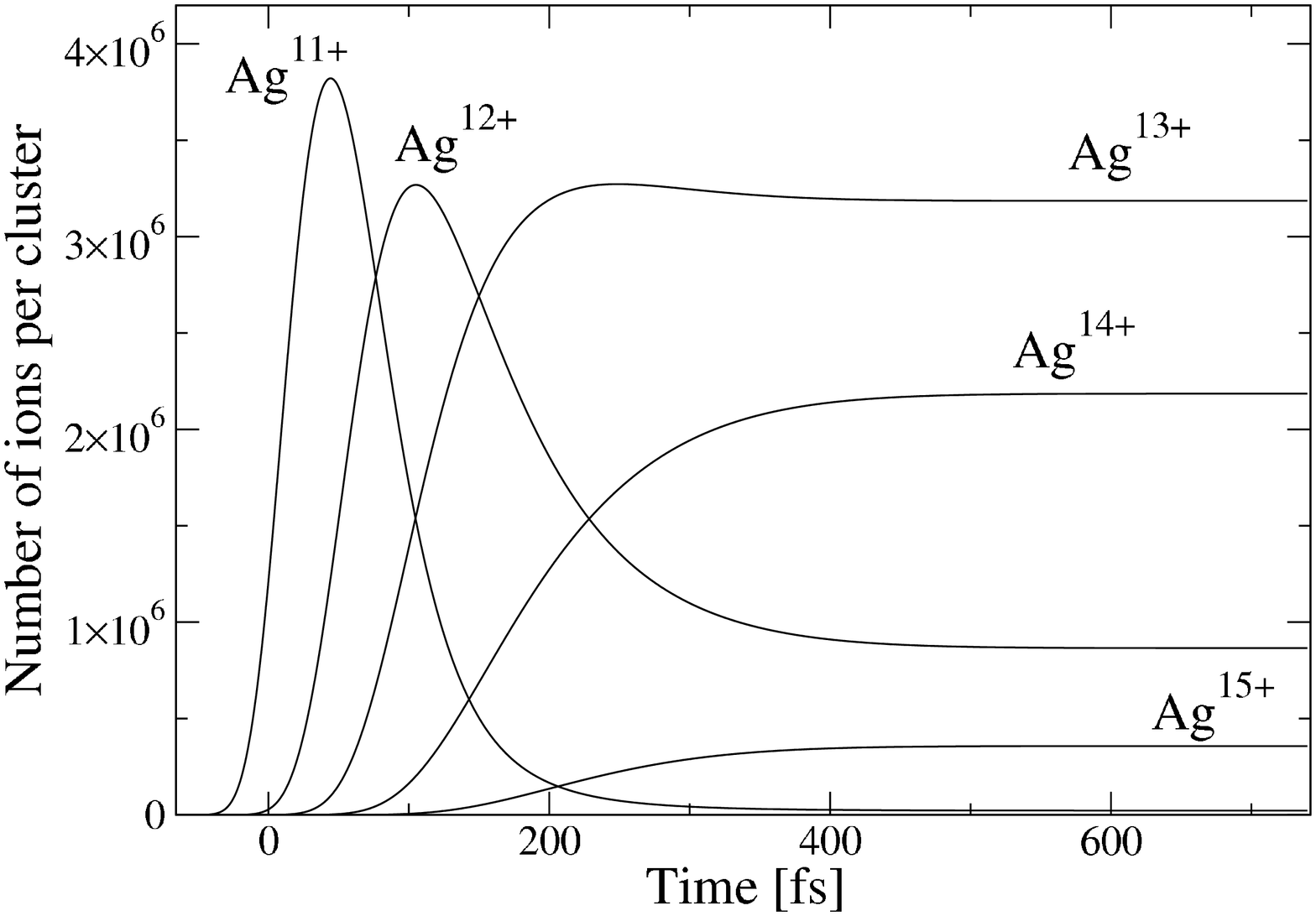}
\caption{{Composition as a function of time for a silver nanoplasma
with initial cluster diameter of 60nm. Lowering of ionization
energies with Stewart-Pyatt shift is included. The pulse parameters
are the same as in Fig.~\ref{comp_silber_single_o_shift}.
\label{comp_silber_single_w_shift}}}
\end{minipage}
\end{figure}
%###################################################################################
Electron temperature and density in a silver cluster with an initial
diameter of 60 nm  are shown in Figs.~\ref{temp_silber_single} and
~\ref{dichte_silber_single} as a function of time. Three different
peak intensities of 80 TW${\rm cm}^{-2}$, 500 TW${\rm cm}^{-2}$, and 3
PW${\rm cm}^{-2}$ are considered. Temperature and density rise
quickly at the beginning of the interaction of the laser pulse with
the cluster. The step-like behavior of the electron density
reflects the inner sequential ionization of the ions in different
charge states. For a pulse with higher intensity, temperature and
density are higher, as expected. For long times, there is a stronger
decrease of density and temperature for pulses with higher
intensities which is a consequence of faster expansion.

An important result are the consequences of the lowering of the
ionization energies which we included in our model using generalized
(nonideal) impact ionization rates. The inclusion of this
nonideality effect (solid lines) leads to lower electron
temperatures whereas the electron densities are higher compared to
the results using the usual ideal rates (dashed lines). Due to the
lowering of the ionization energies, the number of ionization events
in the dense nonideal nanoplasma is higher.

The consequences on the population dynamics for the different ionic
charge states are shown in Figs.~\ref{comp_silber_single_o_shift}
and \ref{comp_silber_single_w_shift}. A silver cluster is considered
with an initial diameter of 60 nm irradiated by a pulse with a peak
intensity of 3 PW${\rm cm}^{-2}$. The comparison shows that the
inclusion of the lowering of the ionization energies in the model
leads to the population of considerable higher ionic charge states
compared to the results using the usual ideal ionization rates. As
can be seen from Fig.~\ref{comp_silber_single_w_shift}, mainly Ag$^{11+}$
and Ag$^{12+}$ are populated during the laser pulse with 130 fs (FWHM).
For later times, i.e. in the expansion
phase after the pulse, Ag$^{13+}$ and Ag$^{14+}$ are the dominating
ionic charge states whereas Ag$^{10+}$ and Ag$^{11+}$ are the
dominating species using the usual ideal rates (see
Fig.~\ref{comp_silber_single_o_shift}).
%###################################################################################
\begin{figure}[htb]
\begin{minipage}[t]{.45\textwidth}
\includegraphics[width=0.9\textwidth,angle=270]{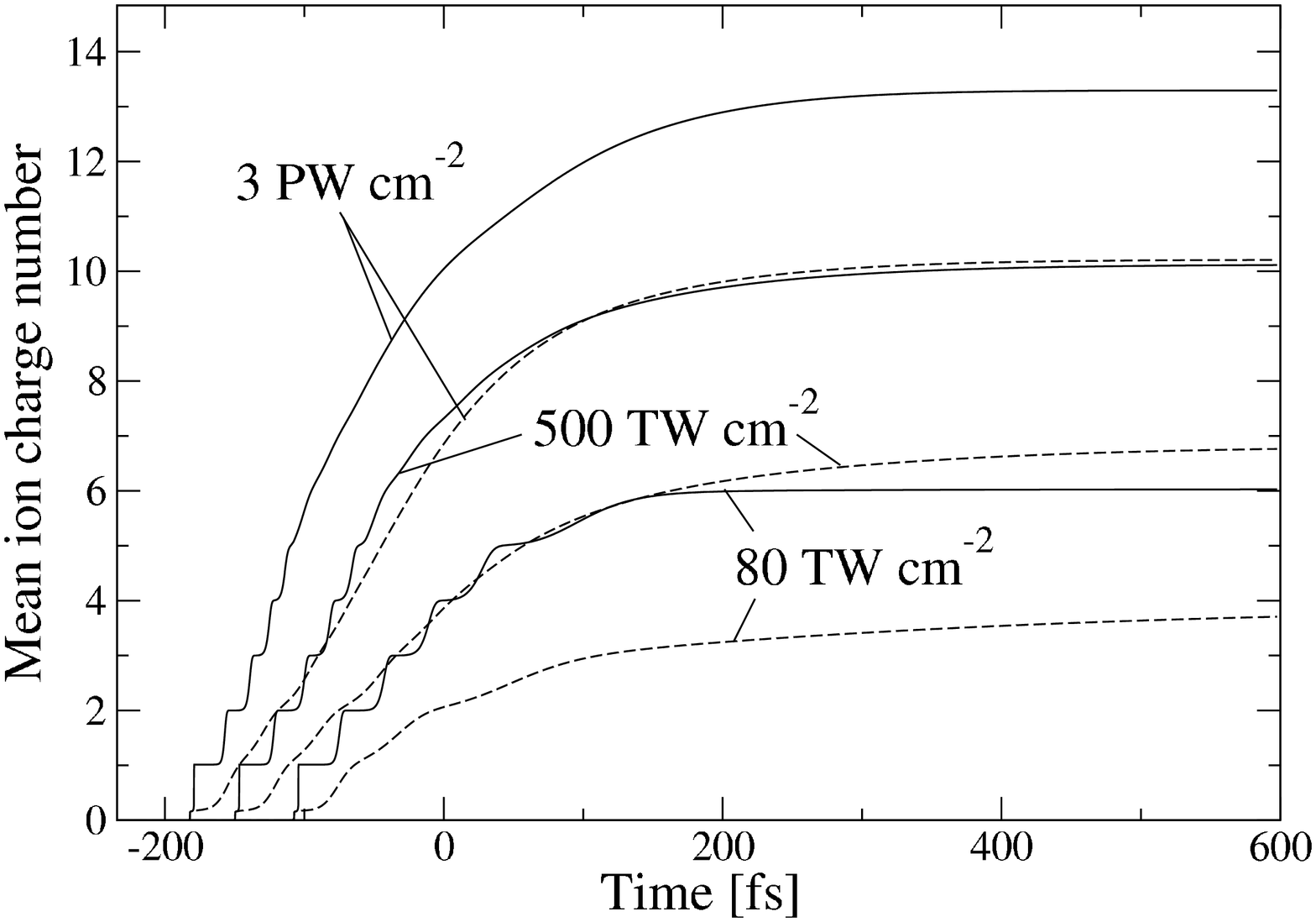}
\caption{{Mean charge number as a function of time for a silver
nanoplasma with initial cluster diameter of 60nm for different
intensities of single 130 fs pulses at 825 nm. Solid line: Lowering
of ionization energies with Stewart-Pyatt shift. Dashed line: No
shift. \label{charge_silber_single}}}
\end{minipage}
\hfil
\begin{minipage}[t]{.45\textwidth}
\includegraphics[width=0.9\textwidth,angle=270]{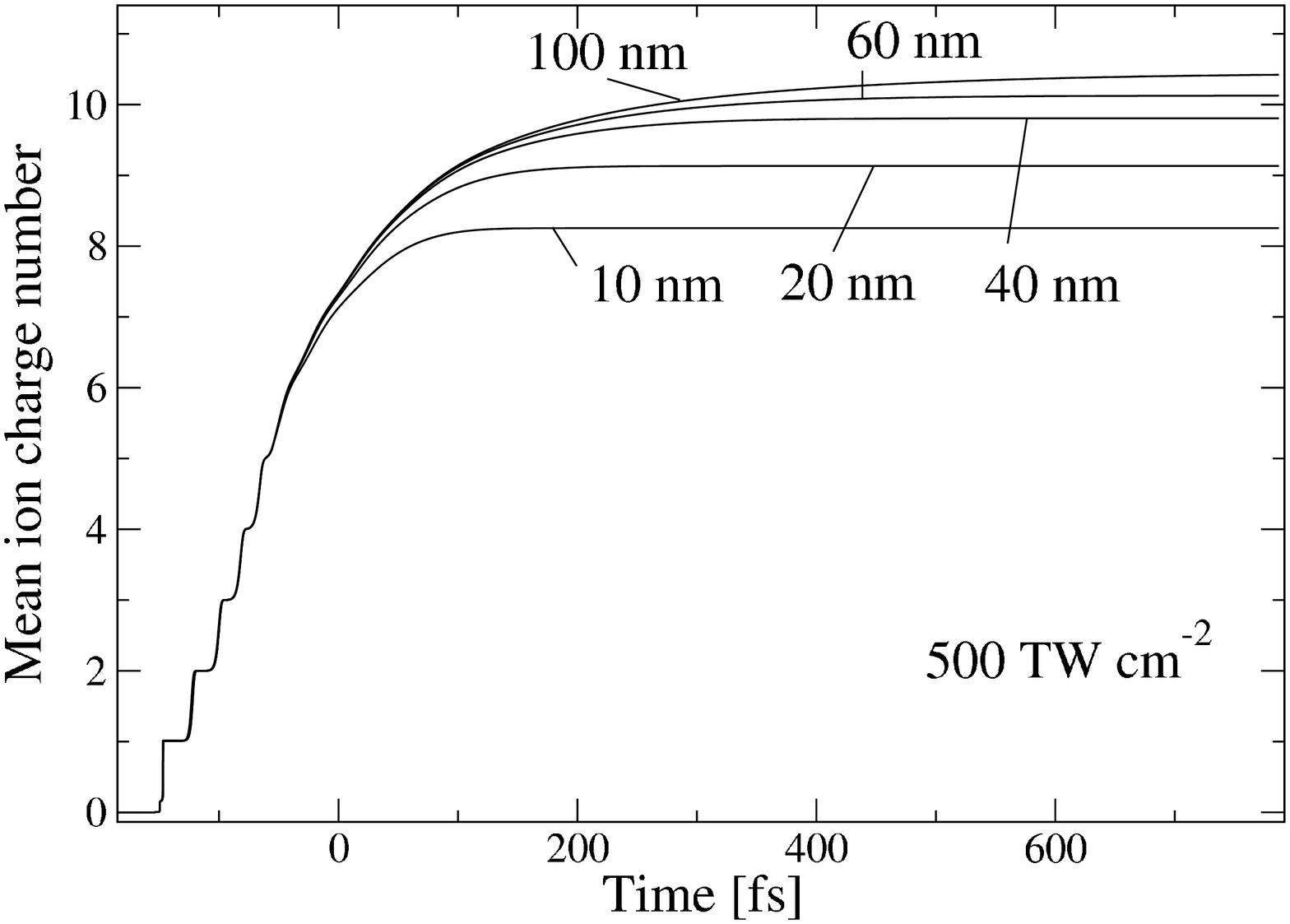}
\caption{{Mean charge number as a function of time for silver
clusters with different initial cluster diameters for 500 TW${\rm
cm}^{-2}$ peak intensity of single 130 fs pulses at 825 nm.
\label{charge_silber_diameter}}}
\end{minipage}
\end{figure}
%###################################################################################

The mean ionic charge number vs. time is shown in
Fig.~\ref{charge_silber_single} for a silver cluster with an initial
diameter of 60 nm.  Again the two variants of the calculation (with
and without lowered ionization energies) are compared for the three
different peak intensities of 80 TW${\rm cm}^{-2}$, 500 TW${\rm
cm}^{-2}$, and 3 PW${\rm cm}^{-2}$. As in
Fig.~\ref{dichte_silber_single}, the step-like behavior at the
beginning of the laser-cluster interaction reflects the ionization
of the ionic species in the different charge states. The higher the
intensity the earlier the mean ion charge number begins to increase
up to a constant value for longer times after the pulse. For the
higher intensities, the mean charge state is increased by a value of about three
compared to that following from the calculations using ideal
rates (dashed).
%###################################################################################
\begin{figure}[htb]
\begin{minipage}[t]{.45\textwidth}
\includegraphics[width=0.9\textwidth,angle=270]{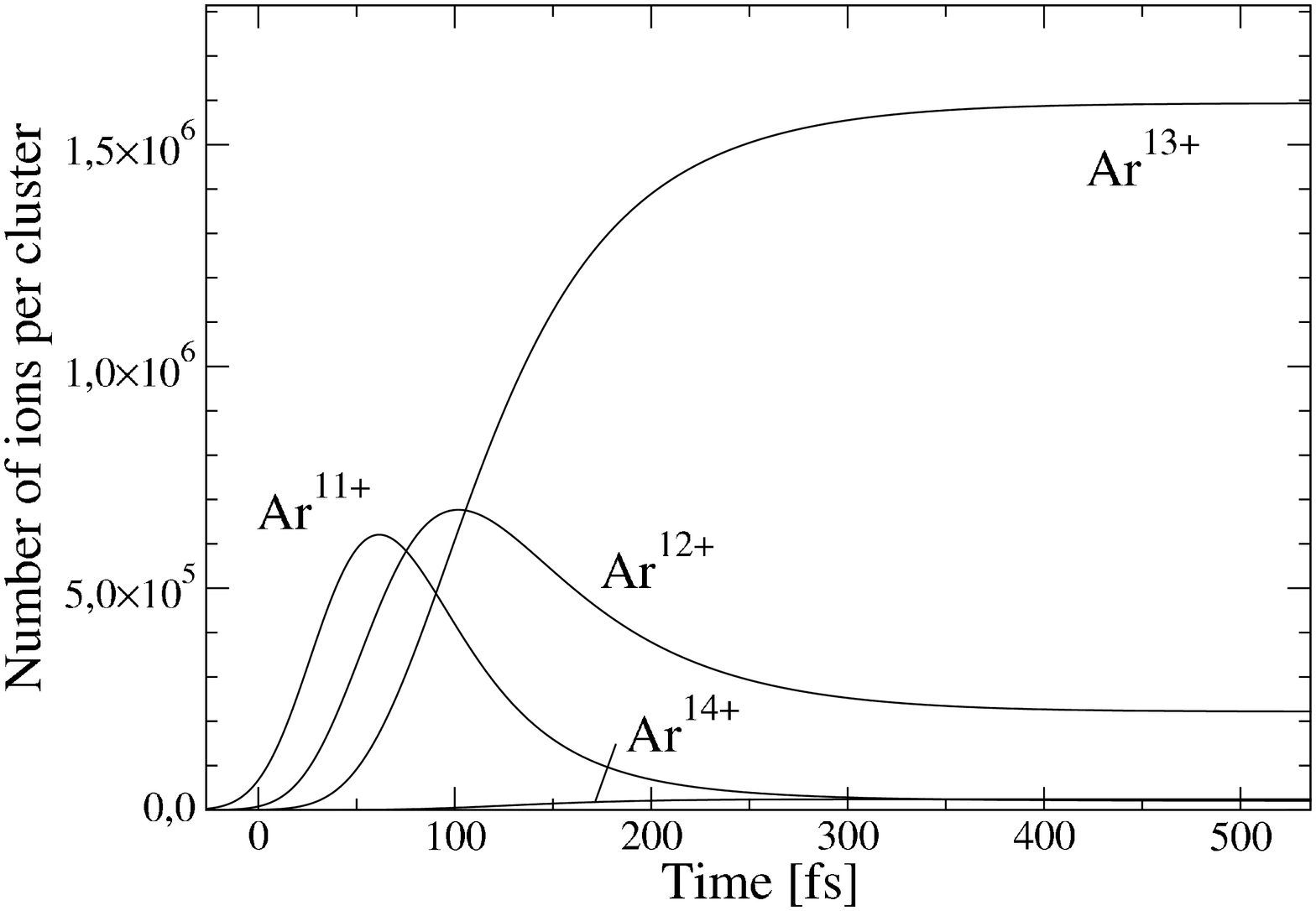}
\caption{{Composition as a function of time for an argon nanoplasma
with initial cluster diameter of 60nm. There was no shift of
ionization energies considered in this calculation. The laser pulse
is Gaussian with 130fs length and a peak intensity of 10PW${\rm
cm}^{-2}$. \label{comp_argon_single_o_shift}}}
\end{minipage}
\hfil
\begin{minipage}[t]{.45\textwidth}
\includegraphics[width=0.9\textwidth,angle=270]{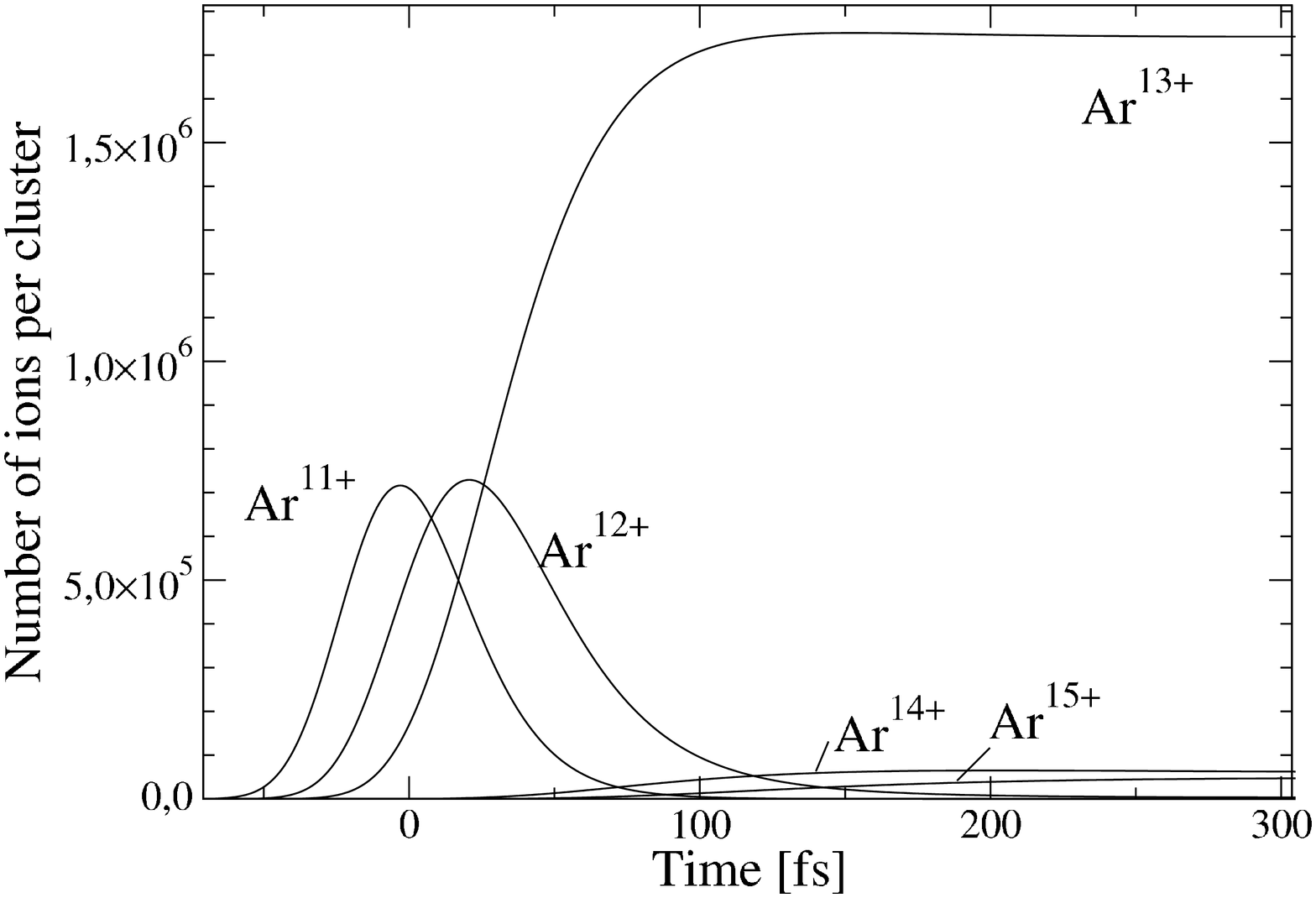}
\caption{{Composition as a function of time for an argon nanoplasma
with initial cluster diameter of 60nm. Lowering of ionization
potentials with Stewart-Pyatt shift is included. The pulse
parameters are the same as in Fig.~\ref{comp_argon_single_o_shift}.
\label{comp_argon_single_w_shift}}}
\end{minipage}
\end{figure}
%###################################################################################

An interesting question is how the mean ionic charge number
calculated within our model depends on the cluster size. Results
concerning this question are presented in
Fig.~\ref{charge_silber_diameter}. In particular, the mean charge is
shown as a function of time for clusters with different initial
diameters irradiated by a laser pulse with peak intensity of 500
TW${\rm cm}^{-2}$. In the early stage of the laser-cluster
interaction a step-like increase of the mean charge number is
observed without remarkable differences between the results for the
different cluster sizes. Essential deviations appear after the
maximum of the pulse: with increasing cluster size, the mean ionic
charge number increases, i.e. for larger clusters the impact
ionization is more efficient during the cluster expansion compared
to smaller clusters.

Let us now investigate the ionization dynamics in a rare gas
cluster. As an example, we consider an argon cluster with an initial
diameter of 60 nm irradiated by a laser pulse with 130 fs length
(FWHM) and peak intensity of 10 PW${\rm cm}^{-2}$. Again, we compare
the results for the population dynamics of the different ionic
charge states following from calculations without and with inclusion
of the lowering of the ionization energies. There are not such
drastic changes concerning the dominating charge states as compared
to silver clusters what results from the differences in the energy
spectra. For long times, there are relatively small quantitative
differences for the population of the dominating charge state
Ar$^{13+}$.

%###################################################################################
\begin{figure}[htb]
\begin{minipage}[t]{.45\textwidth}
\includegraphics[width=0.9\textwidth,angle=270]{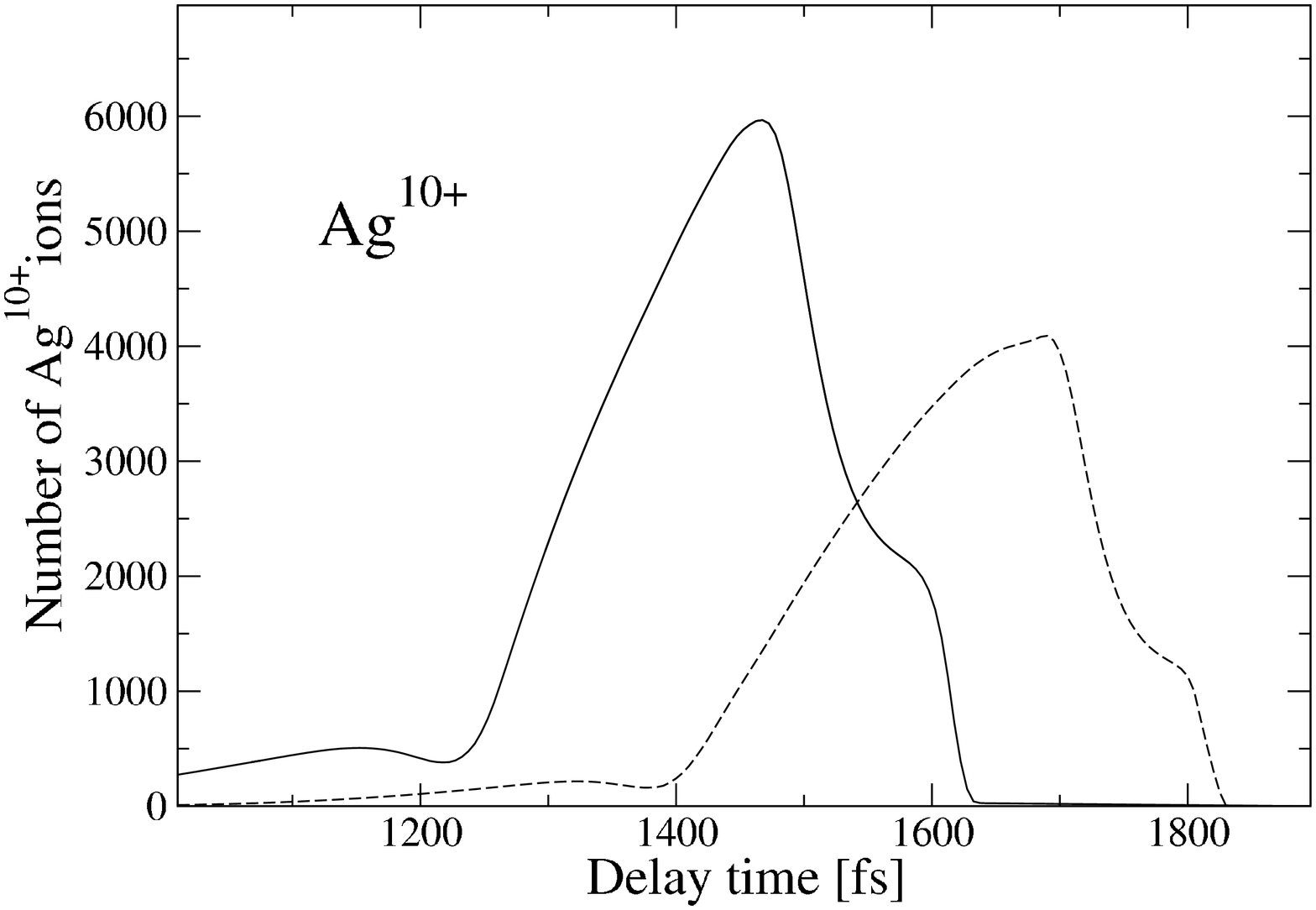}
\caption{{Yield of Ag$^{10+}$ for silver clusters with an initial
diameter of 10 nm irradiated by a double pulse as a function of the
delay time. The pulses have a peak intensity of 80 TW${\rm cm}^{-2}$
and a duration of 130 fs (FWHM). The wavelength is 825 nm. Solid
line: Lowering of the ionization energies with the Stewart-Pyatt
shift. Dashed line: no shift. \label{double_1}}}
\end{minipage}
\hfil
\begin{minipage}[t]{.45\textwidth}
\includegraphics[width=0.9\textwidth,angle=270]{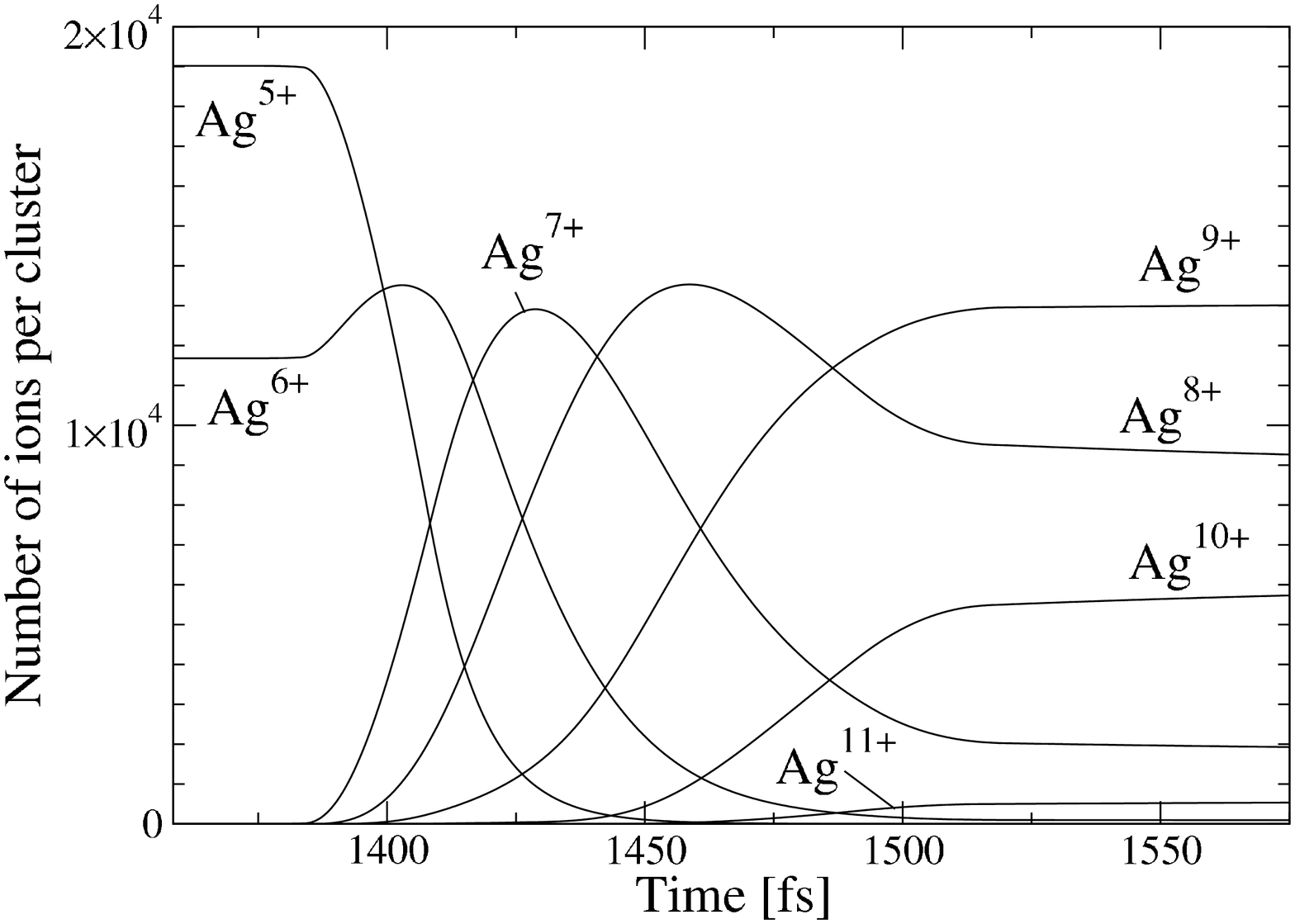}
\caption{{Composition in a silver nanoplasma as a function of time
for a double pulse with a delay time of 1467 fs. The initial cluster
diameter is 10 nm, and the pulse parameters are the same as in
Fig.~\ref{double_1} (with shift). \label{double_2}}}
\end{minipage}
\end{figure}
%###################################################################################

Finally, we present results for double pulse excitation of silver
clusters as investigated recently in experiments by D{\"o}ppner et
al. \cite{doeppner06}. In our calculations, the laser radiation was
simulated by two pulses with Gaussian shape in the intensity profile
with a duration of 130 fs (FWHM) and a peak intensity of 80
TW cm${^{-2}}$ for each pulse. The wavelength was 825 nm. Numerical
results for the yield of Ag$^{10+}$ ions as a function of the delay
time for a silver cluster with an initial diameter of 10 nm are
shown in Fig.~\ref{double_1}. We observe a well-developed increase
of the yield of Ag$^{10+}$ near a certain optimum delay time with a
maximum at a delay of 1467 fs. To explain this, one has to examine
the dynamics of the dense nanoplasma during the laser irradiation
more in detail. The first pulse reaches its maximum at t=0 fs. It
creates, via tunnel ionization and, later,  thermal ionization, a
large overcritical electron density. After the first pulse, the
cluster expands and the electron density decreases. The heating of
the second pulse depends strongly on the applied delay time.
Choosing the delay of the second pulse in such a way that it hits
the cluster at about 3n$_{\rm crit}$ one excites the cluster resonantly
which leads to the Ag$^{10+}$ yield shown in Fig.~\ref{double_1}.
This is in qualitative agreement with the experimental results
where, however, the optimum delay time was found to be of about 3
times larger. To explain this deviation, a more detailed
analysis of the experimental conditions is necessary. For instance,
a cluster size distribution was used in the considered experiment in
contrast to the calculations where a single cluster size was
assumed. In our calculations the maximum Ag$^{10+}$ ion yield is
higher, and it is shifted to a shorter optimum delay if the lowering
of the ionization energy is included (solid line). A more
complicated structure of the Ag$^{10+}$ ion yield as a function of
the delay time is found for larger clusters which will be discussed
elsewhere.

Fig.~\ref{double_2} shows the calculated population of the different
ionic charge stats as a function of time with a delay of 1467 fs for
the considered cluster. The first pulse creates silver ions with
Ag$^{5+}$ and Ag$^{6+}$ being the dominant species. The second pulse
with the same intensity and duration creates considerable higher
ionic charge states with Ag$^{10+}$ being one of the dominant
species.

%######################################################################
\section{Conclusion}
We investigated theoretically the ionization dynamics in silver
clusters and also in argon clusters irradiated by intense laser
pulses within the nanoplasma model. Single pulse and double pulse
excitations were considered applying pulses with a Gaussian shape in
the intensity profile. The calculations were performed for pulses
with the same length of 130 fs (FWHM), but for different peak
intensities. Clusters with initial diameters between 10 nm and 100
nm were considered. The dynamics of the laser-cluster interaction
for single pulse excitation was studied more in detail for clusters
with an initial diameter of 60 nm.

Special attention was paid to the consequences of the inclusion of
the lowering of the ionization energies on the dynamics of the
laser-cluster interaction within the nanoplasma model. In the regime
of the dense nonideal nanoplasma, the bound state properties can be
strongly affected by the interaction with the surrounding particles.
An important effect is the ionization energy suppression which was
treated in the present paper in the Stewart-Pyatt approach. The
change of the reaction rates due to the lowering of the ionization
enegies was taken into account by a simple analytical formula which
allows a qualitative correct description.

The comparison of calculations with the usual ideal rates with the
nonideal rates shows significant changes in the ionization dynamics
for silver clusters. Considerable higher ionic charge states occur
to be the dominant species if nonideal rates are used in the model.
Smaller changes are observed for argon clusters due to the
differences in the energy spectra.

Motivated by recent experiments, the dynamics of laser-cluster
interaction for double pulse excitation of silver clusters was
studied within the present model at the end of the paper. A
well-developed increase of the yield of Ag$^{10+}$ ions near a
certain optimum delay time between the two pulses was obtained which
is in qualitative agreement with the experimental results. At the
optimum delay the second pulse excites the cluster resonantly at
about 3n$_{\rm crit}$ which leads to an increased yield of highly
charged ions.

\section*{Acknowledgments}
This work was supported by the Deutsche Forschungsgemeinschaft (SFB 652). 
We acknowledge valuable discussions with T. D{\"o}ppner, T. Fennel, 
K.-H. Meiwes-Broer, and J. Tiggesb{\"a}umker.

\end{document}